\documentclass[sigconf, screen, anonymous=false]{acmart}
\usepackage{tabularx}
\usepackage{colortbl}
\usepackage{soul}
\usepackage{environ}
\usepackage{mdframed}
\usepackage{hyperref}
\usepackage{multirow}
\usepackage{float}
\usepackage{ragged2e}
\usepackage{xpatch}
\usepackage{microtype}
\usepackage{balance}
\usepackage{tcolorbox}
\usepackage{makecell}
\usepackage{graphicx}
\usepackage{caption}
\usepackage{subcaption}
\usepackage{xurl}
\usepackage[table]{xcolor}
\usepackage{graphicx} 
\usepackage{array}
\usepackage{subcaption}
\usepackage{soul}
\usepackage{needspace}

\setcopyright{none} 
\renewcommand\footnotetextcopyrightpermission[1]{}

\definecolor{headergray}{HTML}{EFEFEF}

\makeatletter
\xpatchcmd{\endmdframed}
  {\aftergroup\endmdf@trivlist\color@endgroup}
  {\endmdf@trivlist\color@endgroup\@doendpe}
  {}{}
\makeatother

\newcommand{\RevisionColor}{black}

\begin{document}

\newcounter{FindingsCounter}
\NewEnviron{finding}[1][]{%
	\vspace{0.1cm}
	\refstepcounter{FindingsCounter}
	
    \begin{mdframed}[backgroundcolor=white]
        \textbf{Finding \arabic{FindingsCounter}}: \BODY
    \end{mdframed}
    \vspace{0.2cm}
}

\NewEnviron{researchquestion}[1][]{%
	\vspace{0.2cm}
    \begin{mdframed}[backgroundcolor=white]
        \BODY
    \end{mdframed}
    \vspace{0.2cm}
}

\title{A Comparison of Conversational Models and Humans in Answering Technical Questions: the Firefox Case}

\author{João Correia}
\affiliation{%
  \institution{Pontifical Catholic University}
  \city{Rio de Janeiro}
  \state{RJ}
  \country{Brazil}
  }
\email{jcorreia@inf.puc-rio.br}

\author{Daniel Coutinho}
\affiliation{%
  \institution{Pontifical Catholic University}  
  \city{Rio de Janeiro}
  \state{RJ}
  \country{Brazil}
  }
\email{dcoutinho@inf.puc-rio.br}

\author{Marco Castelluccio}
\affiliation{%
  \institution{Mozilla Corporation}
  \city{London}
  \state{}
  \country{UK}
  }
\email{mcastelluccio@mozilla.com}

\author{Caio Barbosa}
\affiliation{%
  \institution{Pontifical Catholic University}  
  \city{Rio de Janeiro}
  \state{RJ}
  \country{Brazil}
  }
\email{csilva@inf.puc-rio.br}

\author{Rafael de Mello}
\affiliation{%
  \institution{Federal University of Rio de Janeiro}  
  \city{Rio de Janeiro}
  \state{RJ}
  \country{Brazil}
  }
\email{rafaelmello@ic.ufrj.br}

\author{Anita Sarma}
\affiliation{%
  \institution{Oregon State University}  
  \city{Corvallis}
  \state{OR}
  \country{USA}
  }
\email{anita.sarma@oregonstate.edu}

\author{Alessandro Garcia}
\affiliation{%
  \institution{Pontifical Catholic University}  
  \city{Rio de Janeiro}
  \state{RJ}
  \country{Brazil}
  }
\email{afgarcia@inf.puc-rio.br}

\author{Marco Gerosa}
\affiliation{%
  \institution{Northern Arizona University}
  \city{Flagstaff}
  \state{AZ}
  \country{USA}
  }
\email{marco.gerosa@nau.edu}

\author{Igor Steinmacher}
\affiliation{%
  \institution{Northern Arizona University}
  \city{Flagstaff}
  \state{AZ}
  \country{USA}
  }
\email{igor.steinmacher@nau.edu}

\renewcommand{\shortauthors}{Correia et al.}


\begin{abstract}
The use of Large Language Models (LLMs) to support tasks in software development has steadily increased over recent years. From assisting developers in coding activities to providing conversational agents that answer newcomers’ questions. In collaboration with the Mozilla Foundation, this study evaluates the effectiveness of Retrieval-Augmented Generation (RAG) in assisting developers within the Mozilla Firefox project. We conducted an empirical analysis comparing responses from human developers, a standard GPT model, and a GPT model enhanced with RAG, using real queries from Mozilla’s developer chat rooms. To ensure a rigorous evaluation, Mozilla experts assessed the responses based on helpfulness, comprehensiveness, and conciseness. The results show that RAG-assisted responses were more comprehensive than human developers (62.50\% to 54.17\%) and almost as helpful (75.00\% to 79.17\%), suggesting RAG’s potential to enhance developer assistance. However, the RAG responses were not as concise and often verbose. The results show the potential to apply RAG-based tools to Open Source Software (OSS) to minimize the load to core maintainers without losing answer quality. Toning down retrieval mechanisms and making responses even shorter in the future would enhance developer assistance in massive projects like Mozilla Firefox.


\end{abstract}



\maketitle

\vspace{-1px}
\section{Introduction} 
\label{sec:intro}

\newcommand{\IntroChanges}{
{\color{\RevisionColor} 
The integration of conversational agents into software engineering workflows is transforming how developers interact with code and documentation. Large Language Models (LLMs) tools, such as GitHub Copilot and ChatGPT, are now widely used to support code comprehension, code completion, and debugging \cite{dakhel2023github, sallou2023breaking, wang2024rocks, pinto2023large}. These tools have the potential to significantly reduce the time spent searching online documentation, waiting for human guidance, or interpreting error messages \cite{guo2024exploring}.

Despite their advantages, the utility of general-purpose LLMs can be limited when addressing the specific context of an individual software project~\cite{fan2023large}. Models available to the public are usually trained on a broad range of data. Consequently, their responses may lack context, be outdated, incomplete, or even misleading, especially when dealing with large and actively evolving systems.

To address these limitations, Retrieval-Augmented Generation (RAG) has emerged as a promising approach. RAG architectures enrich language models with retrieval mechanisms that dynamically fetch relevant information, such as project documentation, code snippets, or archived conversations, from curated repositories before generating a response. This grounding enables the model to produce more accurate and context-sensitive answers tailored to a target project \cite{lewis2020retrieval, hao2025RAG}.

Such contextualized answers, tailored to a specific task, are particularly important in Open-Source Software (OSS) projects that depend on a large set of contributors, many of whom are newcomers learning on their own \cite{FagerholmOnboarding2014, steinmacher2014barriers}, to sustain project growth. Newcomers to OSS projects often struggle with the complexity of the software, the diversity of contributions, and the lack of available mentors \cite{steinmacher2015social, steinmacher2021being, feng2022case, FagerholmOnboarding2014}. Although experienced contributors try to support newcomers via pull request reviews or chat interactions, such personalized support may not be sustainable as the project grows. It can divert experienced developers, already overwhelmed \cite{guizani5199260community}, from their efforts to make core contributions \cite{feng2022case, steinmacher2021being}.


RAG-based assistants, while not a replacement to human mentors, are a promising solution to help newcomers overcome project-specific barriers, but implementing and maintaining scalable RAG systems is resource-intensive, requiring substantial engineering effort for data curation, indexing, and retrieval integration. While \citet{correia2024unveiling} took an important first step by comparing RAG-generated answers to human responses, their study did not assess whether such augmentation is necessary. Specifically, they did not include a baseline comparison with general-purpose LLMs (e.g., GPT). Moreover, their evaluation relied on a limited dataset, a single expert assessor, and a narrow, one-dimensional satisfaction metric.
As a result, we still lack a multidimensional understanding of how general-purpose and RAG-based models differ in practice, what trade-offs they present, and whether the added complexity of RAG delivers proportional benefits in large OSS ecosystems.



We conducted a comparative study that analyzed answers provided by different assistants to real technical questions within the context of the Mozilla Firefox project. We introduced a retrieval-augmented generation (RAG) approach that integrates project-specific knowledge to deliver timely, relevant support to developers. Our study compared this RAG system with two other sources: (i) GPT, a general language model assistant, and (ii) Human answers from the Firefox developer community. We evaluated answers to 52 questions, incorporating assessments from Mozilla engineers based on four key attributes: helpfulness, comprehensiveness, conciseness, and overall preference. This expert evaluation offers a more nuanced understanding of each assistant’s strengths and limitations, providing a realistic assessment of their effectiveness in practical development scenarios.

} 
} 

\IntroChanges

Our study reveals that \textit{RAG}-assisted answers were more comprehensive than human responses (62.5\% vs. 54.1\%) while maintaining a comparable level of helpfulness (75.0\% vs. 79.1\%). However, \textit{RAG} responses were generally more verbose and not as concise as human answers. Additionally, \textit{RAG} responses were selected as the preferred option in practice more often than those from \textit{GPT} alone (39.5\% vs. 25.6\%), highlighting the benefits of integrating project-specific knowledge into AI-driven developer assistance. While \textit{RAG} responses offered more contextually relevant and informative answers, conciseness remains an area for improvement.

Thus, this paper presents the following contributions: (i) we evaluate the effectiveness of \textit{RAG} in assisting developers within the Mozilla Firefox project, comparing it to human responses and a standard \textit{GPT} model; (ii) we analyze the trade-offs between helpfulness, comprehensiveness, and conciseness in human, \textit{GPT}, and \textit{RAG} responses, demonstrating that \textit{RAG} provides more detailed and informative answers while maintaining high helpfulness scores; (iii) we identify the strengths and limitations of \textit{RAG}-based approaches for developer assistance, showing that while \textit{RAG} enhances comprehensiveness, response verbosity remains a challenge, and; (iv) we provide insights into the practical application of AI-assisted developer support, emphasizing how \textit{RAG}-based models can alleviate the burden on maintainers in large-scale OSS projects.

\vspace{-5px}
\section{Related Work} \label{sec:background}
{\color{\RevisionColor}
Recent research has explored how large language models can assist with software development tasks and help new developers onboard on existing projects.
}

\vspace{-5px}
\subsection{LLMs in Software Engineering}

\newcommand{\RelatedSoftwareChanges}{
{\color{\RevisionColor}
LLM tools have become increasingly integrated into software development workflows, with recent empirical studies revealing diverse application patterns beyond simple code generation \cite{mathews2024test}. Khojah et al. \cite{khojah2024beyond} investigated how professional engineers utilize ChatGPT during work tasks, finding that developers primarily leverage LLMs for conceptual guidance rather than direct code generation, although concerns about hallucinations and legal limitations affect their adoption. Nam et al. \cite{nam2024using} demonstrated that LLM-based code understanding tools, when integrated directly into IDEs, can significantly improve task completion rates compared to traditional web search, with benefits varying by experience level. Recent advances have addressed repository-level challenges, with CodePlan \cite{bairi2024codeplan} introducing planning algorithms that enable LLMs to perform complex multi-file edits, and ClarifyGPT \cite{mu2024clarifygpt} improving code generation through interactive requirement clarification. Studies of human-AI interaction patterns \cite{nguyen2024beginning} show that beginning programmers struggle with the full prompt-and-edit cycle even for problems at their skill level. 
}
}
\RelatedSoftwareChanges

\vspace{-5px}
\subsection{Supporting Newcomers to Software Projects}

\newcommand{\RelatedNewcomersChanges}{
{\color{\RevisionColor}
Research on newcomer support has evolved from identifying barriers \cite{steinmacher2015social,steinmacher2014barriers} to developing technological interventions that address both technical and social integration challenges. Steinmacher et al. \cite{steinmacher2016overcoming} developed an information portal that provides newcomers with project-specific guidance, finding that structured information delivery significantly reduces initial barriers to contribution. Traditional support mechanisms have included mentoring programs \cite{canfora2012going,steinmacher2021being,feng2025multifaceted}, documentation improvements \cite{treude2011programmers,fronchetti2023contributing}, and social integration strategies \cite{dagenais2010moving} that emphasize human-to-human knowledge transfer. Early technological approaches focused on recommendation systems for finding suitable tasks \cite{xiao2022recommending,santos2021can} and potential mentors \cite{steinmacher2012recommending}. However, these approaches often struggle with scalability limitations, outdated information, interaction barriers, and the substantial effort required from experienced team members.

Recent work has also explored LLM applications for newcomer support, with Azanza et al. \cite{azanza2024can} emphasizing the need for self-service resources that reduce mentor workload, while Balfroid et al. \cite{balfroid2024towards} found that automated code tour generation shows promise but suffers from quality issues, including redundancy and overly detailed explanations. Richards et al. \cite{richards2024what} proposed ToMMY, a code understanding assistant that leverages Theory of Mind (ToM) to infer the mental state of a developer and personalize responses. Emerging AI-assisted approaches \cite{adejumo2024towards} propose using LLMs to address documentation complexity and information overload. However, delivering accurate, contextually relevant assistance at scale remains challenging for off-the-shelf LLMs, which often lack access to up-to-date or project-specific information and may produce generic or hallucinated responses.
}
}
\RelatedNewcomersChanges

\vspace{-5px}
\subsection{Improving LLM Performance with RAG}
{\color{\RevisionColor}
Retrieval-Augmented Generation (RAG) is a promising approach to enhance LLM effectiveness with application in different domains, such as customer support \cite{Xu_2024} and software engineering \cite{ibtashamreqrag, finsaas2024optimizing}. RAG combines the strengths of information retrieval and language generation by allowing LLMs to access and incorporate external sources, such as project documentation or source code, thereby grounding their responses in accurate, question-relevant content rather than relying solely on pre-trained knowledge.

Recent research demonstrates RAG's effectiveness in software engineering contexts. Ibtasham et al. \cite{ibtashamreqrag} investigated the implementation of RAG in software release management through ReqRAG, a specialized chatbot. Their industry-based study revealed that RAG-based responses achieved 70\% relevance and usefulness ratings. Similarly, Finsas and Maksim \cite{finsaas2024optimizing} explored RAG for technical support through relevance feedback and multi-agent interactions. Their research demonstrated that incorporating relevance feedback mechanisms significantly improved response accuracy. However, the potential for RAG to facilitate newcomer onboarding in software projects remains largely unexplored.  
} 

\newcommand{\RelatedComparisionChanges}{
{\color{\RevisionColor}
To the best of our knowledge, \citet{correia2024unveiling} is the only previous work to systematically compare RAG-enhanced LLM responses with those provided by experienced developers for newcomer questions in OSS. Their study introduced DevMentorAI, evaluated using 14 technical questions sourced from a small OSS project, PDF.js. One expert conducted a blind assessment, comparing human-generated and RAG-generated responses. DevMentorAI's answers were considered more satisfactory in 8 cases, equally satisfactory in 3, and less satisfactory in 3. While their work evidences the potential of using RAG-enhanced LLMs for supporting newcomers to OSS, its evaluation was limited, and the assessments were relatively small-scale. Besides, the RAG system used in \citet{correia2024unveiling} was developed in-house, presenting a significant technical barrier to adoption across different projects. Our study significantly extends \citet{correia2024unveiling}'s work in multiple aspects:
(i) We provide a more refined understanding of the usefulness of different answers, with four evaluation attributes that are critical to answers to developer questions \cite{kou2023automated}—helpfulness, comprehensiveness, conciseness, and overall preference—providing a deeper understanding of each approach's strengths and weaknesses, whereas \citet{correia2024unveiling} relied on a single binary generic criterion. Our evaluation methodology includes both individual assessment of standalone attributes and comparative analysis, providing a more robust assessment of answer quality.
(ii) By including GPT, a readily available model, we better isolate RAG's contribution through direct comparison of RAG-enhanced LLM, general-purpose LLM, and human responses. This three-way comparison enables more precise identification of RAG's unique advantages for technical developer inquiries.
(iii) We expanded the scale and methodological rigor by involving 8 experienced engineers rather than a single evaluator, assessed 52 real development questions versus 14, conducted agreement and consistency analyses to mitigate individual biases, and our target project, Firefox, represents a larger, more widely-used open source project compared to PDF.js's smaller repository and knowledge base.
(iv) We adapted Cognita, an open source RAG framework, contrasting with the in-house RAG solution employed in the prior study, thereby ensuring greater reproducibility and applicability. The open source framework was modified to support the ingestion of diverse artifacts from a large-scale project and include a more comprehensive set of knowledge sources across multiple formats.
In summary, our research provides a more comprehensive and methodologically rigorous comparison of human-generated, GPT-generated, and RAG-enhanced GPT-generated answers. By leveraging project-specific artifacts through an open source RAG framework, engaging multiple expert evaluators, and significantly expanding the question set, we offer deeper insights into the practical effectiveness of these systems for supporting newcomer developers with technical questions.

} 
} 
\RelatedComparisionChanges

\vspace{-5px}
\section{Study Design}
\label{subsec:studydesign}

This section outlines the research questions and the methodology employed to address them. Our target project is Mozilla Firefox, a large-scale open source project. We collected technical questions from three Firefox developers' chat rooms hosted on Matrix~\cite{Matrix}. We also collected the human answers for these questions directly in these chat rooms. We generated answers to the same questions using two automated approaches: GPT-4o (\textit{GPT}) alone and GPT-4o enhanced with (\textit{RAG})---leveraging relevant Mozilla Firefox project artifacts as contextual information. Then, Mozilla Firefox engineers assessed the quality of the answers. Their feedback was analyzed quantitatively and qualitatively. In this section, we further detail each of the steps of our study.

\vspace{-5px}
\subsection{Research Questions} \label{subsec:rqs}

Our study is structured around the following research questions:

\vspace{1mm} \noindent
$RQ_1$: \textit{How do experts perceive answers from Humans, GPT, and RAG provided to real technical questions?} 

\newcommand{\ResearchTypoChange}{
{\color{\RevisionColor} We hypothesize that the perceived quality of answers---measured by helpfulness, comprehensiveness, and conciseness \cite{kou2023automated}—}} \ResearchTypoChange will differ due to the inherent characteristics of each source. Firstly, Human responses are expected to embed nuanced, context-specific insights drawn from real-world experience, while \textit{GPT} responses rely on general pre-trained knowledge that may not fully capture project-specific details. In contrast, \textit{RAG} answers by incorporating retrieval mechanisms, could offer a more context-aware perspective. Concluding that these differences significantly impact expert perceptions may imply that AI-based approaches require tailored enhancements to match the practical value of human-provided assistance. In this RQ, we also aim to understand whether experts’ evaluations vary not only in overall quality but also in the specific attributes of helpfulness, comprehensiveness, and conciseness, building on previous findings that underscore the importance of contextual relevance in technical communication.

\vspace{1mm} \noindent
\newcommand{\ResearchQuestionChanges}{
{\color{\RevisionColor}$RQ_2$: \textit{How do helpfulness, comprehensiveness, and conciseness influence experts’ preferences for answers?}
} 
} 
\ResearchQuestionChanges

Given that the effectiveness of a technical answer is often determined by its specific attributes, this research question is driven by the need to understand which characteristics most strongly influence expert preferences. We seek to systematically classify the attributes of responses—specifically helpfulness, comprehensiveness, and conciseness—provided by human developers, \textit{GPT}, and \textit{RAG}. In doing so, we aim to determine whether differences in these attributes, such as the depth of context or the brevity of explanations, impact the overall quality of an answer. Additionally, we intend to comprehend how these attributes interplay and whether certain strengths in one attribute might compensate for weaknesses in another. By answering this research question, we aim to enrich our understanding of how the intrinsic qualities of technical answers affect their practical applicability in real-world software development, ultimately guiding improvements in AI-assisted support tools.

\vspace{1mm} \noindent
$RQ_3$: \textit{Which are the positive and negative aspects of answers from Humans, GPT, and RAG?} 

Since technical answers can exhibit a range of qualitative strengths and weaknesses, this research question is driven by the need to understand the positive and negative aspects of responses provided by human experts, \textit{GPT}, and \textit{RAG}. We seek to conduct a systematic qualitative analysis to classify the benefits and drawbacks of each approach. In particular, we aim to identify aspects such as clarity, contextual relevance, and technical accuracy as positives, while also exploring limitations like verbosity, potential inaccuracies, and a lack of nuanced understanding as negatives. However, it is important to note that qualitative analysis inherently cannot capture every subtle facet of answer quality---such as the evolving perceptions of individual experts or the contextual dependencies that emerge in real-world interactions. By examining these qualitative dimensions, we aspire to provide actionable insights into how each type of response supports or hinders effective communication in technical discussions, ultimately informing the refinement of AI-based assistance tools and guiding future research to enhance their practical usability in software development environments.

\vspace{-5px}
\subsection{Data Collection}
\label{sub:data_coll}

\begin{figure*}[ht!]
  \centering
  \includegraphics[width=0.8\linewidth]{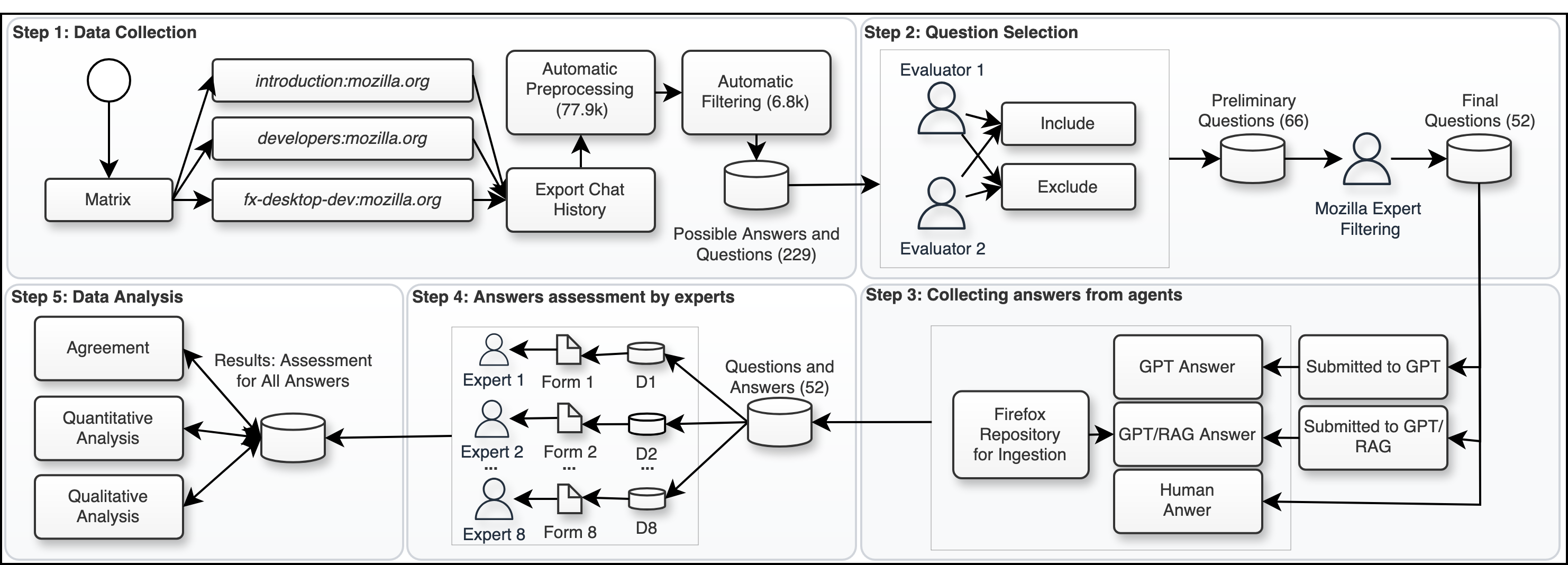}
  
  \vspace{-10px}
  \caption{Study design (The numbers in parentheses refer to the volume of data in each stage of the method).}
  \label{fig:study-design}
  \vspace{-10px}
\end{figure*}

\subsubsection{Step 1: Data Collection}

All the data analyzed is publicly available and was downloaded using the platform feature. The Mozilla community utilizes Matrix.org for communication and question answering, which motivated our choice to analyze chats hosted on this platform. To begin our data collection, we explored all Mozilla-related public channels available on Matrix.org. Subsequently, one of the authors—a Mozilla Firefox developer—selected key channels explicitly focused on Firefox. Ultimately, we collected data from three specific Mozilla Matrix.org channels: \textit{introduction:mozilla.org}, \textit{developers:mozilla.org}, and \textit{fx-desktop-dev:mozilla.org}.



For each selected channel, we exported a raw file containing the complete chat history, including messages, events, and attachments. The chat histories were exported in \textit{JSON} format using the platform's \textit{Export Chat History} feature. 
Considering the fragmented nature of the chat conversations and the difficulty in identifying questions and answers, we applied preprocessing using Python over each chat history. 

\newcommand{\DesignMessageChanges}{
The preprocessing involved the following steps:
  {\color{\RevisionColor}
  \begin{enumerate}
      \item \textit{Message Aggregation}: Sequential messages from the same author were merged into a single entry, preserving the timestamp of the first message. 
      
      \item \textit{Text Normalization}: Replacement of double quotes with single quotes to maintain consistency, removal of excessive whitespace, standardization of line breaks, and conversion to UTF-8 encoding to handle special characters.
      
      \item \textit{Message Categorization}: Regular expressions were used to classify each message into one of four categories.
    
      \begin{itemize} 
        \item \textbf{Reply}. Messages containing at least one line beginning with ">" indicating an explicit reference to a previous message.
        \item \textbf{Question}. Messages containing one or more question marks (?) signaling interrogative intent.
        \item \textbf{Reply to Question}. Messages where a line begins with ">" and terminates with "?" indicating direct responses to earlier questions.
        \item \textbf{General Purpose}. Messages not matching any of the above patterns or consisting solely of whitespace.
      \end{itemize}
      
      \item \textit{Manual Inspections and Adjustments}: Manual inspection and corrections were performed to ensure data quality and accuracy of the preprocessing results.
 \end{enumerate}

} 
}
\DesignMessageChanges

\vspace{-5px}
\subsubsection{Step 2: Question Filtering}

\newcommand{\QuestionFilteringChanges}{
{\color{\RevisionColor} 


We compiled the processed messages into a structured spreadsheet containing Message Type, Timestamp, Author, and Message Content, resulting in approximately 77,900 messages. We then applied temporal filtering to focus on messages from 2024 (the year in which we conducted the study), reducing the dataset to 6,895 messages. Next, we filtered for messages categorized as \textit{Reply to Question}, which yielded 229 messages for 2024. These messages were manually reviewed to ensure they contained both a clear question and a corresponding human answer, with no exclusions made during this review process. This final dataset of 229 question-answer pairs represents 3.3\% of all 2024 messages and 10\% of all \textit{Reply to Question} messages in the complete dataset.
}
}
\QuestionFilteringChanges

\vspace{-5px}
\paragraph{Expert Filtering.}


\newcommand{\ExpertFilteringChanges}{
{\color{\RevisionColor}

After compiling the initial set of possible \textit{Reply to Questions} messages, individually, two evaluators (\emph{Evaluator 1} and \emph{Evaluator 2} in Figure ~\ref{fig:study-design}) manually reviewed the messages to determine which ones would be included or excluded from the study. The evaluators excluded questions that had one or more of the following characteristics: unrelated to Firefox; not addressing source code, documentation, features, or contributions; addressing follow-up of another question; lacking sufficient context for meaningful evaluation of experts or AI models; time-sensitive questions (e.g., version-specific queries); messages that were not questions.

After the individual assessments, the evaluators engaged in an in-person discussion to reach a consensus and resolve differences regarding 15 questions. A question was only included in the study dataset if both evaluators agreed to it. Ultimately, we generated a total of 66 preliminary questions. After this process, a developer from the Firefox project conducted a final review, leading to a refined set of \emph{Final Questions (52)} to be used in the study (22.7\% of the initial dataset of 229 questions). The expert provided the following reasons for excluding additional questions: questions with answers that redirect the user to another channel or user, questions unrelated to Firefox that the first two evaluators had not identified, and questions that were misleading or irrelevant. The questions and the exclusion criteria are available in the supplementary material.}} 
\ExpertFilteringChanges

\vspace{-5px}
\subsubsection{Step 3: Collecting Answers from Models}

{\color{\RevisionColor}
Our goal is to compare responses from three sources: original human answers (Human), GPT-4o-generated responses (GPT), and GPT-4o responses enhanced with Retrieval-Augmented Generation (RAG). Human answers were obtained directly from chat messages, while GPT and RAG responses were generated using the prompt specification described below.} 
\newcommand{\OneParagrah}{
{\color{\RevisionColor}
This prompt was refined through initial testing; early GPT responses tended to be overly detailed and verbose, which is unsuitable for the concise nature of chat environments. To address this, we added a directive limiting responses to a single paragraph, ensuring consistency and practical applicability across all three answer types.}
}\OneParagrah

\vspace{-0.5em}
\begin{tcolorbox}[colback=white, colframe=black!50, coltitle=black, colbacktitle=headergray, fonttitle=\bfseries, title=LLM Prompt to Collect the Answers]
\small
\vspace{-2mm}The question below was taken from a Mozilla Firefox developer's chat. Answer the question directly and in one paragraph, as long replies may disturb the chat.

Question: <question>.\vspace{-2mm}
\end{tcolorbox}
\vspace{-0.5em}



To collect answers using \textit{RAG}, we utilized an open source Framework named Cognita~\cite{Cognita}. Initially, we adapted and extended the embedding components within Cognita to support the ingestion of file formats that were originally unsupported, such as source code files. These improvements allowed us to successfully ingest technical documentation and source code from our target project, i.e., \texttt{Firefox}.
\newcommand{\SourceArtifactsRagChange}{
{\color{\RevisionColor}
All artifacts were ingested exclusively from the publicly available Gecko-Dev GitHub repository (\url{https://github.com/mozilla/gecko-dev}).
}}\SourceArtifactsRagChange
Notably, our modified component supports the ingestion of source code in multiple programming languages and markup formats. Specifically, we ingested a variety of artifacts, including source code, wiki pages, and web documents in formats such as \textit{js}, \textit{jsx}, \textit{mjs}, \textit{cjs}, \textit{md}, \textit{html}, \textit{pdf}, among others. \newcommand{\VerificationIngestion}{
{ \color{\RevisionColor}
To verify the effectiveness of the artifact ingestion process, we ran a set of representative queries covering key Firefox development topics and manually inspected the top retrieved chunks to confirm that the system consistently returned semantically relevant and accurate information from various known artifact types (e.g., source code, documentation). All the tests demonstrated that the ingestion and retrieval components reliably grounded the RAG model in project-specific knowledge.
}}\VerificationIngestion

\newcommand{\DesignPromptChanges}{
{ \color{\RevisionColor}
Using the Cognita Framework for RAG purposes, we enhanced the previously mentioned prompt by incorporating a specific context section. This section includes relevant information retrieved from project artifacts based on the semantic similarity between the user's question and the ingested content, which includes source code comments, documentation, and wiki pages (sources suggested by the Firefox developer). The framework embeds the user query and searches for the most semantically similar chunks using a vector database, which are then appended to the prompt to support context-aware generation. We submitted queries manually to both GPT and RAG using their web interfaces, and the resulting responses were systematically organized into spreadsheets for analysis. A complete list of the RAG's answers, along with the contextual data retrieved for each, is available in our supplementary material \cite{supmaterial}. Below, we summarize the approaches we used to collect answers from the different sources.

\begin{itemize}
  \item \textbf{Human answer}: extracted directly from the chat history.
  \item \textbf{GPT answer}: extracted using the large language model (GPT-4o) through the web platform \href{http://chat.openai.com}{chat.openai.com}.
  \item \textbf{RAG answer}: extracted using an extended version of the Cognita Framework, through a locally deployed web platform augmented with artifacts from the Firefox repository.
\end{itemize}
} 
} 
\DesignPromptChanges

\newcommand{\CollectionCostChange}{
{\color{\RevisionColor}
The primary cost stemmed from generating embeddings for the large Firefox codebase, which amounted to around USD 28 and was completed in approximately 24 hours. The study required only two model inferences per question, and each inference took less than 5 seconds to execute. The total cost for inferences was USD 3.
}
}\CollectionCostChange

\vspace{-5px}
\subsubsection{Step 4: Expert Assessment of Answers}

A panel of eight Mozilla engineers (referred to as \emph{Expert 1} through \emph{Expert 8} in Figure~\ref{fig:study-design}) evaluated the answers. Each expert received a form with 10 questions with corresponding answers generated by \textit{GPT}, \textit{RAG}, and \textit{Humans}. From these, we randomly assigned six uniquely ones to each expert. The remaining subset of four questions was assigned to all experts. We intentionally included the overlapped evaluations to measure agreement among the eight experts during their assessments.

Before the evaluation, the experts received detailed instructions outlining the evaluation procedure. Specifically, they received clear definitions of the criteria used to assess the sets of answers given to the technical questions. For each question, experts reviewed multiple alternative answers and evaluated them according to three primary attributes: 



\begin{itemize}

\item \textbf{Helpfulness}: The degree to which the answer provides meaningful assistance in addressing the question.
\item \textbf{Comprehensiveness}: The extent to which the answer includes all essential elements required to respond thoroughly to the question.
\item \textbf{Conciseness}: The quality of conveying all necessary information succinctly, without including unnecessary or redundant words.

\end{itemize}


The evaluators followed a structured method to assess each question, allowing for a thorough and unbiased comparison of multiple answers.\newcommand{\DesingQuestionsAnonimChanges}{
{\color{\RevisionColor}To reduce potential bias, all questions and answers were anonymized and presented in a random order. }} \DesingQuestionsAnonimChanges The assessment process was divided into three main phases, as detailed in Table~\ref{tab:answer_evaluation}.

\newcommand{\DesignScaleChanges}{
{\color{\RevisionColor}
In the first phase, each answer was evaluated individually on a binary scale based on three predefined attributes (e.g., helpful or not helpful). This scale was defined in consultation with a Mozilla manager to reduce workload and allow for a more comprehensive evaluation. This approach ensured that answers were evaluated on their own merits before any direct comparisons were made. While a Likert scale could have offered finer-grained feedback, the binary method allowed for clear distinctions among sources and was supplemented by comparative selections in the next phase (e.g., identifying "the most helpful" answer). This supported both absolute and relative quality assessments without adding unnecessary complexity.
} 
} 
\DesignScaleChanges

\vspace{-5px}
\begin{table}[ht!]
\centering
\caption{Illustrative form for the assessment of a question and its answers.} \label{tab:answer_evaluation}
\vspace{-10px}

\resizebox{\columnwidth}{!}{
\begin{tabular}{l|l}
\Xhline{2pt}
\textbf{Question X:} & \textit{<Question content>}. \\
\Xhline{2pt}
& \textit{Multiple Choice} \\
\hline
\textbf{Answer 1:} \textit{<Answer content>} & \newline
[ ] Helpful \quad [ ] Comprehensive \quad [ ] Concise \\
\hline
\textbf{Answer 2:} \textit{<Answer content>} & \newline
[ ] Helpful \quad [ ] Comprehensive \quad [ ] Concise \\
\hline
\textbf{Answer 3:} \textit{<Answer content>} & \newline
[ ] Helpful \quad [ ] Comprehensive \quad [ ] Concise \\
\Xhline{2pt}
& \textit{Single Choice} \\
\hline
\textbf{Most helpful information:} & [ ] Answer 1 [ ] Answer 2 [ ] Answer 3 \\
\textbf{Most comprehensive information:} & [ ] Answer 1 [ ] Answer 2 [ ] Answer 3 \\
\textbf{Most concise without being incomplete:} & [ ] Answer 1 [ ] Answer 2 [ ] Answer 3 \\
\textbf{Preferred answer in practice:} & [ ] Answer 1 [ ] Answer 2 [ ] Answer 3 \\
\Xhline{2pt}
\textbf{Open feedback} & \makecell[l]{Please highlight any particularly relevant or  \\irrelevant aspects of the preferred answer(s), \\or note any missing elements.} \\
\Xhline{2pt}
\end{tabular}
}
\end{table}
\vspace{-1em}


In the second phase, the evaluators selected which answer stood out as the most helpful, comprehensive, and concise. It allowed evaluators to compare the three answers provided to each question. In addition, they indicated which answer from the three alternatives they would prefer to see in practice, i.e., which answer they would most likely rely on in a real-world scenario. 

In the third phase, we asked evaluators to provide feedback through an open question. This gave them the chance to highlight particularly strong or weak aspects of the answers, suggest improvements, and note any missing elements that could enhance the answers. This evaluation process aimed to capture objective judgments and subjective insights from experienced professionals. 


Each evaluator followed the same process described above for the ten questions. In the end, they were also asked to provide demographic information, which is presented in Table~\ref{tab:demographics}. The 1\textsuperscript{st} column shows the evaluator’s ID; the 2\textsuperscript{nd} indicates their confidence level in the assessment, measured on a Likert scale; the 3\textsuperscript{rd} reports their years of contribution to Mozilla, also on a Likert scale; \newcommand{\DesignGenderChanges}{{\color{\RevisionColor}the 4\textsuperscript{th} refers to their gender, an open and optional field showing values as reported by experts }} \DesignGenderChanges and  \newcommand{\DesignProficiencyChanges}{the {\color{\RevisionColor}5\textsuperscript{th} column shows their self-rated proficiency in answering questions in Matrix channels, offering context on their familiarity with developer support practices.}} \DesignProficiencyChanges

\begin{table}[ht!]
\caption{Mozilla experts' demographics.}
\label{tab:demographics}
\vspace{-10px}
\resizebox{\columnwidth}{!}{
\begin{tabular}{cccccc}
\hline
\rowcolor[HTML]{F3F3F3} 
\textbf{ID} & \textbf{\begin{tabular}[c]{@{}c@{}}Confidence \end{tabular}} & \textbf{\begin{tabular}[c]{@{}c@{}}Experience\end{tabular}} & \textbf{\begin{tabular}[c]{@{}c@{}}Contributing\\ to Mozilla\end{tabular}} & \textbf{Gender\textsuperscript{*}} & \textbf{\begin{tabular}[c]{@{}c@{}}Q\&A \\ Proficiency\end{tabular}} \\ \hline
\rowcolor[HTML]{FFFFFF} 
{\color[HTML]{434343} E1} & {\color[HTML]{434343} 3} & {\color[HTML]{434343} 10+ years} & {\color[HTML]{434343} 10+ years} & {\color[HTML]{434343} Male} & {\color[HTML]{434343} 5} \\ \hline
\rowcolor[HTML]{FFFFFF} 
{\color[HTML]{434343} E2} & {\color[HTML]{434343} 4} & {\color[HTML]{434343} 4-7 years} & {\color[HTML]{434343} 4-7 years} &  -- & {\color[HTML]{434343} 5} \\ \hline
\rowcolor[HTML]{FFFFFF} 
{\color[HTML]{434343} E3} & {\color[HTML]{434343} 3} & {\color[HTML]{434343} 10+ years} & {\color[HTML]{434343} 1-3 years} & {\color[HTML]{434343} Male} & {\color[HTML]{434343} 3} \\ \hline
\rowcolor[HTML]{FFFFFF} 
{\color[HTML]{434343} E4} & {\color[HTML]{434343} 3} & {\color[HTML]{434343} 10+ years} & {\color[HTML]{434343} 1-3 years} & -- & {\color[HTML]{434343} 4} \\ \hline
\rowcolor[HTML]{FFFFFF} 
{\color[HTML]{434343} E5} & {\color[HTML]{434343} 4} & {\color[HTML]{434343} 10+ years} & {\color[HTML]{434343} 10+ years} & {\color[HTML]{434343} Female} & {\color[HTML]{434343} 5} \\ \hline
\rowcolor[HTML]{FFFFFF} 
{\color[HTML]{434343} E6} & {\color[HTML]{434343} 5} & {\color[HTML]{434343} 8-10 years} & {\color[HTML]{434343} 4-7 years} & {\color[HTML]{434343} Male} & {\color[HTML]{434343} 5} \\ \hline
\rowcolor[HTML]{FFFFFF} 
{\color[HTML]{434343} E7} & {\color[HTML]{434343} 3} & {\color[HTML]{434343} 10+ years} & {\color[HTML]{434343} 10+ years} & -- & {\color[HTML]{434343} 3} \\ \hline
\rowcolor[HTML]{FFFFFF} 
{\color[HTML]{434343} E8} & {\color[HTML]{434343} 4} & {\color[HTML]{434343} 10+ years} & {\color[HTML]{434343} 8-10 years} & {\color[HTML]{434343} Male} & {\color[HTML]{434343} 5} \\ \hline
\end{tabular}
}
\vspace{-10px}
\end{table}

\begingroup
\renewcommand{\thefootnote}{}
\footnotetext{\textsuperscript{*} As reported by the participants}
\endgroup

\vspace{-5px}
\section{Data Analysis}

As described in Section \ref{sub:data_coll}, experts evaluated responses from \textit{Humans}, \textit{GPT}, and \textit{RAG} based on three primary attributes: helpfulness, comprehensiveness, and conciseness. Each expert assessed a subset of questions, with four questions evaluated by all experts to allow inter-rater agreement analysis. \textit{Fleiss’ Kappa} was used to measure agreement among the eight experts on these common questions, yielding kappa values of 0.67 (Substantial), 0.69 (Substantial), 0.31 (Fair), and 0.45 (Moderate). These findings indicate an overall acceptable level of agreement while also revealing variability that may stem from differences in question clarity or expert interpretation. See complete data in our supplementary material \cite{supmaterial}.

To compare expert perceptions of responses from the three sources, we computed the proportion of answers exhibiting each attribute. Additionally, for each question, we identified the answer that best represented each attribute (i.e., the most helpful, the most comprehensive, and the most concise). To determine statistical significance in these variations, we performed pairwise comparisons using Fisher’s Exact Test, evaluating whether differences between sources were significant. The contingency tables used in this analysis are presented in Table~\ref{tab:three_columns}. These tables provide the frequency distributions of assigned attributes across answer sources used for statistical comparisons. This analysis directly addresses $RQ_1$ by quantifying how experts assess answer quality across sources and identifying notable differences in response attributes.

\begin{table}[ht!]
  \centering
  \vspace{-5px}
  \caption{Contingency tables for Fisher's test.}
  \vspace{-10px}

  \resizebox{\columnwidth}{!}{
  \begin{tabular}{lll}
    \hline
    \rowcolor[HTML]{F3F3F3} 
    \multicolumn{1}{c}{\cellcolor[HTML]{F3F3F3}\textbf{Helpfulness}} & \multicolumn{1}{c}{\cellcolor[HTML]{F3F3F3}\textbf{Comprehensiveness}} & \multicolumn{1}{c}{\cellcolor[HTML]{F3F3F3}\textbf{Conciseness}} \\ 
    \hline
    $\begin{array}{l|cc}
      & \textbf{Human} & \textbf{GPT} \\ \hline
      \textbf{Yes} & 38 & 26 \\
      \textbf{No} & 10 & 22 \\ \hline
      \multicolumn{1}{r}{\textbf{Total}} & \multicolumn{2}{c}{48} \\[2mm]
    \end{array}$ 
    &
    $\begin{array}{l|cc}
      & \textbf{Human} & \textbf{GPT} \\ \hline
      \textbf{Yes} & 26 & 19 \\
      \textbf{No} & 22 & 29 \\ \hline
      \multicolumn{1}{r}{\textbf{Total}} & \multicolumn{2}{c}{48} \\[2mm]
    \end{array}$
    &
    $\begin{array}{l|cc}
      & \textbf{Human} & \textbf{GPT} \\ \hline
      \textbf{Yes} & 12 & 14 \\
      \textbf{No} & 36 & 34 \\ \hline
      \multicolumn{1}{r}{\textbf{Total}} & \multicolumn{2}{c}{48} \\[2mm]
    \end{array}$ \\
    \hline
    $\begin{array}{l|cc}
      & \textbf{Human} & \textbf{RAG} \\ \hline
      \textbf{Yes} & 38 & 36 \\
      \textbf{No} & 10 & 12 \\ \hline
      \multicolumn{1}{r}{\textbf{Total}} & \multicolumn{2}{c}{48} \\[2mm]
    \end{array}$ 
    &
    $\begin{array}{l|cc}
      & \textbf{Human} & \textbf{RAG} \\ \hline
      \textbf{Yes} & 26 & 30 \\
      \textbf{No} & 22 & 18 \\ \hline
      \multicolumn{1}{r}{\textbf{Total}} & \multicolumn{2}{c}{48} \\[2mm]
    \end{array}$
    &
    $\begin{array}{l|cc}
      & \textbf{Human} & \textbf{RAG} \\ \hline
      \textbf{Yes} & 12 & 11 \\
      \textbf{No} & 36 & 37 \\ \hline
      \multicolumn{1}{r}{\textbf{Total}} & \multicolumn{2}{c}{48} \\[2mm]
    \end{array}$ \\
    \hline
    $\begin{array}{l|cc}
      & \textbf{GPT} & \textbf{RAG} \\ \hline
      \textbf{Yes} & 26 & 36 \\
      \textbf{No} & 22 & 12 \\ \hline
      \multicolumn{1}{r}{\textbf{Total}} & \multicolumn{2}{c}{48} \\[2mm]
    \end{array}$ 
    &
    $\begin{array}{l|cc}
      & \textbf{GPT} & \textbf{RAG} \\ \hline
      \textbf{Yes} & 19 & 30 \\
      \textbf{No} & 29 & 18 \\ \hline
      \multicolumn{1}{r}{\textbf{Total}} & \multicolumn{2}{c}{48} \\[2mm]
    \end{array}$
    &
    $\begin{array}{l|cc}
      & \textbf{GPT} & \textbf{RAG} \\ \hline
      \textbf{Yes} & 14 & 11 \\
      \textbf{No} & 34 & 37 \\ \hline
      \multicolumn{1}{r}{\textbf{Total}} & \multicolumn{2}{c}{48} \\[2mm]
    \end{array}$ \\
    \hline
  \end{tabular}
  }
  \label{tab:three_columns}
\vspace{-10px}
\end{table}

\newcommand{\CorrDetails}{
{\color{\RevisionColor}
Each assessed question yielded four key indicators: the most helpful answer, the most comprehensive answer, the most concise answer, and the answer the expert would prefer in practice. To analyze the relationship between these indicators and the experts’ preferences, we conducted a non-parametric correlation analysis using Spearman's rank correlation. This method was selected due to the ordinal nature of our data and because it does not assume normal distribution. Spearman’s correlation is suitable for understanding whether higher scores on individual attributes are associated with increased likelihood of an answer being preferred in practice, helping to address $RQ_2$. This allowed us to quantify the relative importance of these attributes in real-world usage scenarios, revealing whether experts prioritize detailed and informative responses over concise but potentially less informative answers.}}
\CorrDetails

\newcommand{\QualiDetails}{
{\color{\RevisionColor}
Additionally, experts provided qualitative feedback on the responses, offering insights into both the strengths and weaknesses of each source. Three researchers manually analyzed this feedback, using an open coding approach. After they reached consensus on the codes, they applied thematic analysis to identify recurring patterns and themes. Codes were first generated inductively from the data and then grouped into broader categories that reflected strengths and weaknesses in each response source.} } \QualiDetails
Through this process, we identified recurring patterns in how each answer source performed under different conditions, including cases where human responses outperformed the models due to domain-specific expertise, and cases where RAG responses excelled in effectively integrating documentation. Similarly, we noted instances where \textit{GPT} responses, despite being concise, lacked necessary depth, leading to lower expert preference.

These qualitative insights form the basis of our assessment for $RQ_3$, helping us contextualize the statistical findings with real-world expert opinions. By combining both quantitative and qualitative analyses, we provide a holistic evaluation of how different answer sources contribute to technical discussions, ultimately guiding future improvements in AI-assisted developer support. 

\vspace{-5px}
\section{Results and Discussions} \label{sec:results}

In this section, we present the results of our study, which aimed to explore answers to real questions related to the Mozilla Firefox project, comparing answers provided by \textit{Humans}, \textit{GPT}, and a \textit{RAG}. We evaluated the answers with the help of Mozilla Firefox engineers, who evaluated each response based on its helpfulness, comprehensiveness, and conciseness. In addition, the evaluators determined the answer that they thought best revealed each of these attributes and picked their preferred answer. 

\vspace{-5px}
\subsection{Overall Performance}
\label{sec:overall_performance}

\newcommand{\ResultsTableChanges}{
{\color{\RevisionColor}
In Table ~\ref{tab:overall-a}, we present the percentage of each attribute assessed per source. The 1\textsuperscript{st} column lists the attributes; the 2\textsuperscript{nd}, the answer source; the 3\textsuperscript{rd}, the percentage of questions where the attribute was assigned; and the 4\textsuperscript{th}, where that source was considered \textit{the most} for that attribute (e.g., \textit{the most comprehensive}). Additionally, Table ~\ref{tab:overall-b} shows the percentage of questions in which each source was preferred for practical use.

\begin{table}[ht!]
\centering
\caption{Answer attribute analysis: assignment, best answer per attribute, and presence in practice.}
\vspace{-5px}

\begin{subtable}[t]{\linewidth}
\centering
\caption{Assigned Attributes and Most Preferred Answer per Attribute.}
\label{tab:overall-a}
\begin{tabular}{l|c|c|c}
\hline
\rowcolor[HTML]{F3F3F3}
\textbf{\small Attribute} & \textbf{\begin{tabular}[c]{@{}c@{}}\small Answer\\ \small Source\end{tabular}} & \textbf{\begin{tabular}[c]{@{}c@{}}\small Assigned\\ \small (\% Questions)\end{tabular}} & \textbf{\begin{tabular}[c]{@{}c@{}}\small The Best \\ \small (\% Questions)\end{tabular}} \\
\hline
\multirow{3}{*}{\small Helpfulness}
 & \small Human & \cellcolor{gray!79}79.1\% & \cellcolor{gray!30}30.2\% \\
 & \small GPT  & \cellcolor{gray!54}54.1\% & \cellcolor{gray!23}23.2\% \\
 & \small RAG  & \cellcolor{gray!75}75.0\% & \cellcolor{gray!46}46.5\% \\
 \cline{2-4}
 & \textit{Total} & --- & \textbf{100.0\%} \\
\hline\hline
\multirow{3}{*}{\small Comprehensiveness}
 & \small Human & \cellcolor{gray!54}54.1\% & \cellcolor{gray!21}20.9\% \\
 & \small GPT  & \cellcolor{gray!39}39.5\% & \cellcolor{gray!21}23.2\% \\
 & \small RAG  & \cellcolor{gray!62}62.5\% & \cellcolor{gray!55}55.8\% \\
 \cline{2-4}
 & \textit{Total} & --- & \textbf{100.0\%} \\
\hline\hline
\multirow{3}{*}{\small Conciseness}
 & \small Human & \cellcolor{gray!25}25.0\% & \cellcolor{gray!36}33.3\% \\
 & \small GPT  & \cellcolor{gray!29}29.1\% & \cellcolor{gray!36}35.7\% \\
 & \small RAG  & \cellcolor{gray!23}22.9\% & \cellcolor{gray!31}30.9\% \\
 \cline{2-4}
 & \textit{Total} & --- & \textbf{100.0\%} \\
\hline
\end{tabular}
\end{subtable}
\hfill
\begin{subtable}[t]{\linewidth}
\caption{Percentage of Answers Preferred to See in Practice.}
\label{tab:overall-b}
\centering
\begin{tabular}{l|c|c|c|c}
\hline
\rowcolor[HTML]{F3F3F3}
 & \textbf{\small Human} & \textbf{\small GPT} & \textbf{\small RAG} & \textbf{\small Total} \\ \hline
 \textbf{\begin{tabular}[c]{@{}l@{}}\small Preferred to See in \\ \small Practice (\% Questions)\end{tabular}} & \cellcolor{gray!35}34.8\% & \cellcolor{gray!26}25.5\% & \cellcolor{gray!40}39.5\% & \textbf{100.0\%} \\
\hline
\end{tabular}
\end{subtable}
\vspace{-10px}
\end{table}

}
}
\ResultsTableChanges

\vspace{-5px}
\subsubsection{Helpfulness}

\textit{Human} answers were assigned this attribute in 79.1\% of the evaluated questions. Closely following were the \textit{RAG} answers with 75.0\% deemed helpful, and \textit{GPT} was rated helpful in only 54.1\% of the cases. When comparing which source produced the \textbf{most helpful} answer, \textit{Human} responses were selected in only 30.2\% of the cases, whereas \textit{RAG} was chosen in 46.5\%, and \textit{GPT} in 23.2\%. This contrast suggests that, while \textit{Human} responses were usually considered helpful, they did not always provide the best assistance, with \textit{RAG} offering actionable responses, and standing out as the most helpful in nearly half of the cases (46.5\%).

\begin{figure}[h]
  \centering
  \begin{subfigure}{0.20\textwidth}
    \centering
    \includegraphics[width=\linewidth]{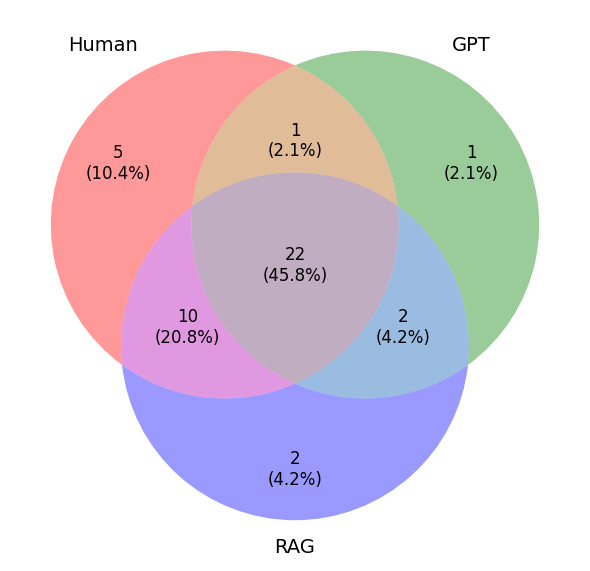}
    \caption{Helpful Answers}
    \label{fig:veen_help}
  \end{subfigure}
  \hspace{0.3cm} 
  \begin{subfigure}{0.20\textwidth}
    \centering
    \includegraphics[width=\linewidth]{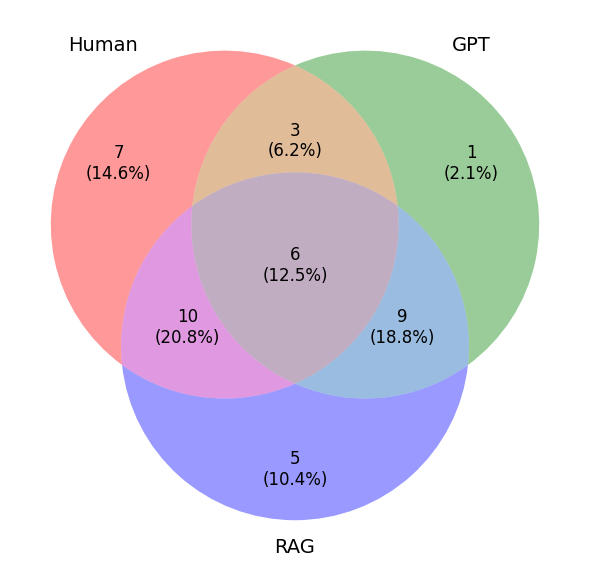}
    \caption{Comprehensive Answers}
    \label{fig:veen_comp}
  \end{subfigure}
  \par\vspace{0.3cm} 
  \begin{subfigure}{0.20\textwidth}
    \centering
    \includegraphics[width=\linewidth]{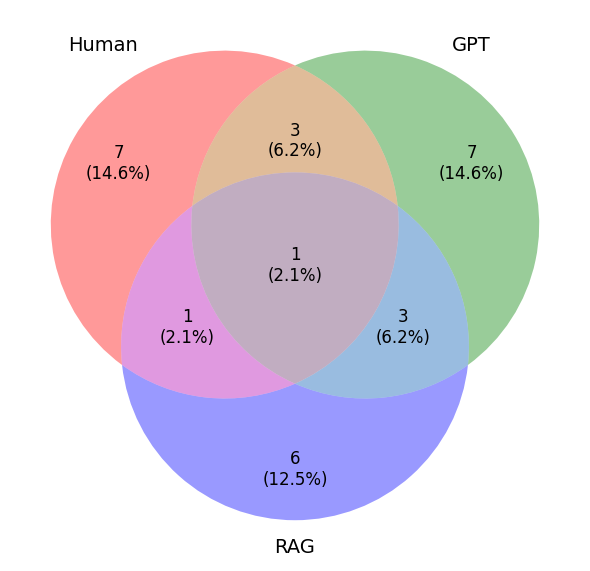}
    \caption{Concise Answers}
    \label{fig:veen_con}
  \end{subfigure}
  \vspace{-5px}
  \caption{Number of answers and the occurrence of attributes}
  \label{fig:venn}
  \vspace{-10px}

\end{figure}

\newcommand{\ResultsVennQualChange}{
{\color{\RevisionColor}
Figure~\ref{fig:venn} provides visual support for qualitative insights into the distribution of helpful, comprehensive, and concise answers across the different sources. }} \ResultsVennQualChange
Focusing on helpfulness, Figure~\ref{fig:veen_help} illustrates that for 22 questions, all three sources provided answers that were considered helpful. \textit{Human} answers account for the highest number of uniquely helpful answers for questions, with 5 exclusive cases, against 2 exclusive cases to \textit{RAG} and 1 to \textit{GPT}. The nearest source to the \textit{Human} (in numbers) is the \textit{RAG}, showing 10 answers for questions deemed as helpful. In contrast, \textit{Human} and \textit{GPT} have only one question where answers from both sources were deemed helpful. It is important to note that, despite the high number of helpful answers attributed to \textit{Human}, \textit{RAG} was the source with more answers rated as \textit{the most helpful} by engineers, reinforcing the impact of retrieval augmentation in generating valuable responses.


\newcommand{\ResultsClarity}{
{\color{\RevisionColor}
Moreover, we observe that RAG provided helpful answers to only two questions that humans themselves did not provide helpful answers to, whereas GPT managed this for only one. This result suggests that, despite RAG's tendency to perform better than GPT in general in terms of helpfulness, the challenge of surpassing humans in this aspect remains for both approaches.
}
}
\ResultsClarity

\vspace{-5px}
\subsubsection{Comprehensiveness}

Concerning \textbf{comprehensiveness}, \textit{RAG} answers were scored as comprehensive for 62.5\% of questions. In contrast, \textit{Human} and \textit{GPT} answers were rated as comprehensive in 54.1\% and 39.5\% of questions, respectively. When assessing which source provided the \textbf{most comprehensive} response, \textit{RAG} was chosen in 55.8\% of cases, followed by \textit{GPT} answers in 23.2\% and \textit{Human} answers in 20.9\%. These results may suggest the \textit{RAG}‘s capacity to retrieve relevant artifacts to construct responses with depth and completeness. In contrast, \textit{GPT} answers without direct access to external references, relying solely on pre-trained knowledge, may have resulted in less detailed responses than \textit{RAG}. Likewise, \textit{Human} answers may have been constrained by time limitations or an assumption that the recipient had some background knowledge, leading to the lowest selection rate for the most comprehensive.

Figure~\ref{fig:veen_comp} highlights the heterogeneity of sources considered as having comprehensive answers. \textit{Human} was the source with more (7 unique cases) comprehensive answers; this may point to the human synthesis capacity. On the other hand, \textit{RAG} follows closely with 5 unique cases. The overlap between \textit{Human} and \textit{RAG} (10 cases) suggests that this source can better approximate \textit{Human} reasoning, in contrast to the overlap of \textit{RAG} and \textit{Human} (3 cases). The \textit{RAG}'s performance (5 unique cases) contributes to the view that LLMs can answer technical questions in OSS. However, it is important to mention that \textit{GPT} alone was comprehensive in only one answer, underscoring its limitations by relying solely on its pre-trained knowledge. The 6 cases in which all sources were equally comprehensive suggest that some questions inherently allow for completeness across different answering methods, at least for well-known software like Mozilla Firefox.

\vspace{-5px}
\subsubsection{Conciseness}
Conciseness is critical in real-time developer discussions, where long answers may lead to a waste of effort. \textit{GPT} answers were considered concise in 29.1\% of questions, while \textit{Human} and \textit{RAG} answers were rated concise in 25.0\% and 22.9\% of questions, respectively. Interestingly, this attribute was the most balanced between the three sources. When selecting the \textbf{most concise} answer, \textit{GPT} was chosen in (35.7\%) of the questions, followed by \textit{Human} in (33.3\%) and \textit{RAG} in (30.9\%). These results may suggest that no source was superior in terms of conciseness. This may suggest that, while participants generally did not find answers concise, they still chose the least verbose option. In this context, \textit{GPT} could be perceived as slightly better than \textit{Human} and the \textit{RAG}, tending to generate direct answers that engineers found to be more concise without losing essential details.

Figure~\ref{fig:veen_con} shows the distribution of conciseness among the answer sources. As described in the last paragraph, conciseness demonstrates a balanced distribution. This suggests that conciseness is not uniformly achievable across different answers for the same question. Notably, \textit{Human} and \textit{GPT} each had 7 unique, concise answers, indicating that both human expertise and language models can independently produce succinct answers. \textit{RAG}, on the other hand, had 6 concise responses, maybe due to the amount of provided information, implying that retrieval augmentation does not necessarily lead to increased conciseness. In addition, there were 3 overlapping cases between \textit{GPT} and \textit{Human}, and 3 between \textit{GPT} and \textit{RAG}, highlighting instances in which conciseness emerged across different methods. Interestingly, in only 1 question, answers from all sources were rated as concise, reinforcing that achieving conciseness is highly dependent on the question and context.

\vspace{-5px}
\subsubsection{Practice}

Ultimately, the evaluators were asked to select the answer they would prefer to see in practice. \textit{RAG} answers were preferred in 39.5\%, followed closely by \textit{Human} answers at 34.8\%, and \textit{GPT} alone at 25.8\%. These percentages are relatively close, suggesting no single approach was overwhelmingly favored. However, it is worth noting that \textit{RAG} had a nearly 15\% and 5\% over \textit{GPT} and \textit{Human}, respectively. Furthermore, it may suggest a preference for responses enriched with retrieval-augmented information and the need to better understand the factors influencing preferences. While \textit{Human} answers were often assigned as helpful and comprehensive, \textit{RAG}'s ability to integrate specific knowledge might have made it a more favorable option in some cases. The fact that human answers were frequently helpful but not always preferred to be seen in practice may suggest that evaluators valued structured, evidence-backed responses over informal or experience-based explanations. The lower preference for \textit{GPT} might indicate that even concise responses could lack the necessary depth required for real-world applications without additional retrieval mechanisms.

\vspace{-5px}
\subsection{Statistical Analysis of Attributes}

We conducted two-sided \textit{Fisher's Exact Tests} to statistically analyze the differences observed among the sources based on the contingency tables (See Table \ref{tab:three_columns}). In Table~\ref{tab:analysis_results}, we summarize the results from the statistical tests, including \textit{p-values} and \textit{odds ratios (OR)}. The \textit{OR} quantifies the strength of association between two groups and an attribute. An \textit{OR} greater than 1 indicates that the first group is more likely to exhibit the attribute, whereas an \textit{OR} less than 1 suggests that the second group is more likely to exhibit it.

\begin{table}[ht!]
  \centering
  \vspace{-5px}
  \caption{Fisher's test results per answer and attribute.}
  \vspace{-10px}
  \resizebox{\columnwidth}{!}{
  \begin{tabular}{lccc}
    \hline
    \rowcolor[HTML]{EFEFEF} 
    \textbf{Comparison} & \textbf{p-value} & \textbf{Odds Ratio} & \textbf{Significance (< 0.05) } \\
    \hline
    \multicolumn{4}{c}{\textbf{Helpfulness}} \\
    \hline
    Human vs GPT & 0.01656 & 3.215 & \textbf{Yes} \\
    Human vs RAG & 0.80858 & 1.267 & \textbf{No} \\
    GPT vs RAG  & 0.05401 & 0.394 & \textbf{No} \\
    \hline
    \multicolumn{4}{c}{\textbf{Comprehensiveness}} \\
    \hline
    Human vs GPT & 0.21959 & 1.804 & \textbf{No} \\
    Human vs RAG & 0.53483 & 0.709 & \textbf{No} \\
    GPT vs RAG  & 0.04063 & 0.393 & \textbf{Yes} \\
    \hline
    \multicolumn{4}{c}{\textbf{Conciseness}} \\
    \hline
    Human vs GPT & 0.81868 & 0.810 & \textbf{No} \\
    Human vs RAG & 1.00000 & 1.121 & \textbf{No} \\
    GPT vs RAG  & 0.64240 & 1.385 & \textbf{No} \\
    \hline
  \end{tabular}
  }
  \label{tab:analysis_results}
  \vspace{-5px}
\end{table}

\vspace{-5px}
\subsubsection{Helpfulness}

The results indicate a significant difference between \textit{Human} and \textit{GPT} responses regarding the helpfulness (\textit{p} = 0.016, OR = 3.215), suggesting that \textit{Human} answers were significantly more likely to be rated as helpful. However, no significant differences were observed between \textit{Human} vs. \textit{RAG} (\textit{p} = 0.808, OR = 1.267), indicating that retrieval augmentation did not meaningfully affect helpfulness compared to human responses.

The comparison between \textit{GPT} and \textit{RAG} (\textit{p} = 0.054, OR = 0.394) is a borderline case. The p-value is slightly above the conventional significance threshold of 0.05, meaning the evidence is not strong enough to reject the null hypothesis, but it suggests a possible trend. The OR indicates that \textit{RAG} responses were more likely to be helpful compared to \textit{GPT}. While this result is not statistically significant under a strict threshold, it does hint at a potential advantage of \textit{RAG} answers that could become clearer with a larger sample size.

\vspace{-5px}
\subsubsection{Comprehensiveness}

There were no significant differences in comprehensiveness ratings between \textit{Human} vs. \textit{GPT} (\textit{p} = 0.219, OR = 1.804) or \textit{Human} vs. \textit{RAG} (\textit{p} = 0.534, OR = 0.709). However, a significant difference was observed between \textit{GPT} and \textit{RAG} (\textit{p} = 0.040, OR = 0.393), indicating that \textit{RAG} responses were significantly more likely to be comprehensive compared to \textit{GPT} alone.

\vspace{-5px}
\subsubsection{Conciseness}

No significant differences were observed in conciseness across the tests, as shown by the \textit{p-value} and OR. It is important to note that not finding a difference does not mean that the responses are equivalent in terms of conciseness.

\vspace{-5px}
\subsubsection{Findings and Discussions}


The study findings demonstrate that \textit{RAG} notably enhances helpfulness and comprehensiveness compared to standard \textit{GPT}, \newcommand{\FindsDifferences}{{\color{\RevisionColor} and performs comparably to \textit{Human} answers across all evaluated attributes. No statistically significant differences were observed between \textit{RAG} and \textit{Human} answers in terms of helpfulness, comprehensiveness, or conciseness.}} \FindsDifferences This suggests that \textit{RAG}-generated responses can match the overall quality of \textit{Human} answers in developer support contexts. 

\vspace{-5px}
\subsection{Attributes Influencing Practical Preference}

\begin{figure}[h]
  \centering
  \includegraphics[width=0.8\columnwidth]{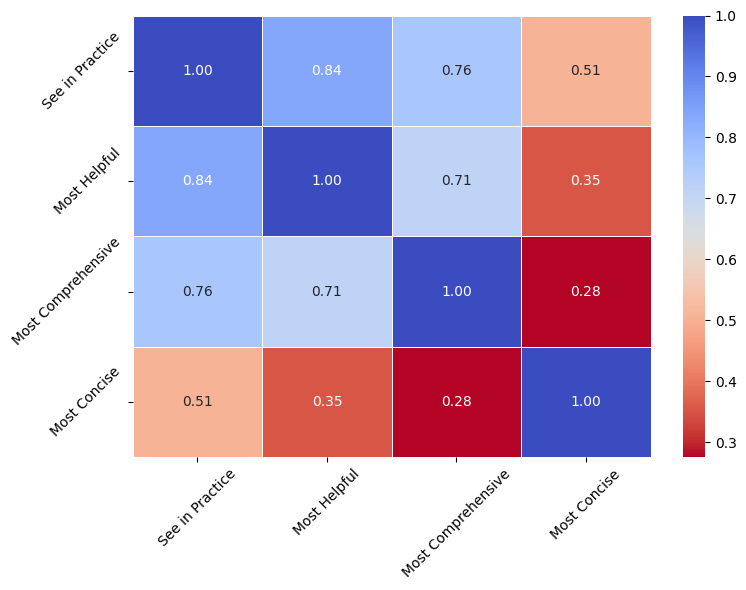}
  \vspace{-10px}
  \caption{Correlation between answers attributes and their likelihood of being seen in practice.}\vspace{-12pt}
  \label{fig:see_in_practice}
\end{figure}

The results described in previous sections provide insights concerning the proportions of answers that experts perceive as helpful, comprehensive, and concise. However, we need to understand the characteristics influencing certain answers to be preferred in real-world scenarios. Thus, we conducted a correlation analysis to determine possible patterns. Figure \ref{fig:see_in_practice} presents the correlation coefficients between the evaluated attributes and the selection of responses as preferred to see in practice. 

The analysis indicates that \textit{helpfulness} shows the strongest correlation (0.84) with \textit{See in Practice}, indicating that responses considered helpful are more likely to be perceived as applicable in real-world scenarios. This finding highlights the implications of providing helpful answers to address the problem effectively.

Next, \textit{comprehensiveness} demonstrates a notable correlation (0.76) with the preference to \textit{See in Practice}, indicating that answers with detailed and well-rounded explanations tend to be more valuable. The ability to integrate documentation and source code into answers probably contributes to this attribute, as it ensures that explanations are substantiated and contextually relevant.

Despite the dynamic context where the content analyzed was obtained, the \textit{conciseness} attribute reached the weakest correlation (0.51) with \textit{see in practice}, suggesting that while brevity is a desirable attribute, it is not a primary concern. Concise responses may lack the necessary depth required for real-world application, potentially limiting their effectiveness despite being easier to read.

\vspace{-5px}
\subsection{Evaluators’ Perspectives}

The evaluators provided detailed insight into the answers, highlighting strengths, weaknesses, and potential practical applications. Their assessments contribute to reaching a nuanced understanding of how \textit{Human}, \textit{GPT}, and \textit{RAG} answers might perform in real-world interactions. In the following, we discuss the main findings.

\vspace{-5px}
\subsubsection{Human Answers}
\label{sub:qua_human}
Human answers were recognized for their correctness and practical relevance. Evaluators consistently highlighted that \textit{Human} answers provided useful information, correctly capturing the intent of the questions and offering solutions grounded in real-world experience. Many responses included direct references to relevant resources, such as links and artifacts, which evaluators found particularly useful (Question 1). Furthermore, we found that \textit{Human} answers frequently selected as the best responses not only address the immediate technical question but also provide additional insights that could help developers solve related challenges.


\vspace{-0.5em}
\begin{mdframed}[frametitle=Question 1, subtitleaboveline=true, frametitlerule=true, subtitlebelowskip=0pt, subtitleaboveskip=2pt, skipabove=\baselineskip, skipbelow=\baselineskip, linecolor=black, linewidth=1.5pt]
  \footnotesize
  I would like to look at some WPT test results. How do I find a recent m-c landed version - or similar - in Treeherder?
  \mdfsubtitle{Human's Answer:}
  \url{https://wpt.fyi/results/?label=experimental&label=master&aligned} is likely easiest. Our results to the dashboard already come from treeherder But if you want logs, you could to to m-c in treeherder and just click on a recent job like \url{https://treeherder.mozilla.org/jobs?repo=mozilla-central&selectedTaskRun=IU6KZJ2uQ6e72CPphRZdGA.0}.
  \mdfsubtitle{Evaluator Comment:}
  I think that having direct links (as long as they are correct links) is always going to be more helpful than plain-word descriptions. 
\end{mdframed} 
\vspace{-0.5em}

Despite their benefits, \textit{Human} answers sometimes lack technical depth or completeness. Some evaluators noted instances where \textit{Human} responses assumed prior knowledge, making them less accessible, for instance, to newcomers. It is also possible that answers were written in a rush, leading to insufficient or less precise guidance. An evaluator explicitly identified this issue, saying that their own \textit{Human} answer was rushed and that the \textit{RAG} answer provided a more useful explanation (Question 2). This scenario emphasizes an important factor in developer interactions: while \textit{Human} responses can be highly effective, time constraints and the informal nature of chat discussions may result in incomplete answers. 

\vspace{-0.5em}
\begin{mdframed}[frametitle=Question 2, subtitleaboveline=true, frametitlerule=true, subtitlebelowskip=0pt, subtitleaboveskip=5pt, skipabove=\baselineskip, skipbelow=\baselineskip, linecolor=black, linewidth=1.5pt]
  \footnotesize
  Is there a helper function somewhere in the frontend to open a new browser window $(BROWSER\_CHROME\_URL)$, and return a promise which resolves when $gBrowser$ is available in that new window?
  \mdfsubtitle{Human's Answer:}
  $BrowserTestUtils$ has some such thing but don't know the name off hand (on the phone sorry).
  \mdfsubtitle{Evaluator Comment:}
  [RAGAnswer] is more useful than [HumanAnswer] which was written on a rush (by me, lol). [GPTAnswer] is wrong, $OpenBrowserWindow$ doesn't return a promise.
\end{mdframed} 
\vspace{-0.5em}

In summary, \textit{Human} answers seem to be appreciated by their accuracy and applicability. However, their too specific focus can sometimes result in gaps, mainly when more detailed explanations or guidance are needed. According to our findings, one possible implication is that RAG responses —with their tendency to offer more extensive explanations —could effectively complement \textit{Human} answers. While \textit{Human} answers remain a gold standard, their overall effectiveness could be further enhanced by integrating more detailed support provided by RAG models.

\vspace{-5px}
\subsubsection{GPT Answers}

The answers generated by \textit{GPT} generally exhibit positive attributes, particularly on conciseness, when compared to those from \textit{RAG} and \textit{Humans}. Evaluators noted that \textit{GPT} answers were well-organized and easy to read, making them especially valuable for developers looking for quick explanations. For instance, when asked how to retrieve a specific version of the Firefox repository using Mercurial, \textit{GPT} provided a clear and direct command without any unnecessary details (Question 3). Evaluators appreciated this focused approach, especially in contrast to other answers that included additional but potentially irrelevant information. While \textit{Human} responses sometimes offered broader insights, such as an alternative to Mercurial, \textit{GPT} answer was often positively evaluated due to their precision and adherence.

\vspace{-0.5em}
\begin{mdframed}[frametitle=Question 3, subtitleaboveline=true, frametitlerule=true, subtitlebelowskip=0pt, subtitleaboveskip=5pt, skipabove=\baselineskip, skipbelow=\baselineskip, linecolor=black, linewidth=1.5pt]
  \footnotesize
  As someone who's new to mercurial (coming from Git), how would I get a version of the Firefox repository, to build FF from source code, as it was (in the main/central branch) on a certain date?
  \mdfsubtitle{GPT's Answer:}
  To get a version of the Firefox repository from a specific date using Mercurial, you can use the following command: `hg update -d "YYYY-MM-DD"`. This will update your local repository to reflect the state of the central branch as it was on the specified date.
  \mdfsubtitle{Evaluator Comment:}
  [GPTAnswer] answers the question without adding the extra (probably unnecessary) information about how to clone the repo like [RAGAnswer] does. [HumanAnswer] is maybe more helpful from a general sense (if the question asker would rather use git than mercurial), but doesn't have any details.
\end{mdframed} 
\vspace{-0.5em}


The \textit{GPT} answers also revealed deficiencies. One primary concern was its tendency to deliver confident yet incorrect answers, which we can describe as plausible content with incorrect information (Question 4). 

\begin{mdframed}[frametitle=Question 4, subtitleaboveline=true, frametitlerule=true, subtitlebelowskip=0pt, subtitleaboveskip=0pt, skipabove=\baselineskip, skipbelow=\baselineskip, linecolor=black, linewidth=1.5pt]
  \footnotesize
  I was able to compile, but the long-time built Firefox is slower than the Firefox produced with artifact build? how can I enable optimization for performance on speedometer 2.1?
  \mdfsubtitle{GPT's Answer:}
  To enable optimization for performance on Speedometer 2.1 in your Firefox build, you should ensure that you're compiling with optimization flags. Use `./mach build --enable-release` to build with optimizations for speed. Additionally, make sure that you are not using debug builds, as they are not optimized for performance.
  \mdfsubtitle{Evaluator comment:}
  The right answer is [HumanAnswer]. [GPTAnswer] just does something that doesn't exist. [RAGAnswer] almost gets it right, but doesn't --enable-release or anything like that. So you really want to look at the shippable mozconfigs in the tree :)
\end{mdframed} 

This confidence in supplying answers could lead users to the wrong directions, especially in cases where the subject matter is complex. We can also note that \textit{GPT} can produce vague responses when a clear answer is unavailable. We might suppose that this issue could be more common in conditions without additional context, but further investigation is required. On the other hand, \textit{GPT} produces coherent responses, but its reliance on pre-trained knowledge without real-time retrieval could make it less reliable for domain-specific questions. Evaluators' comments suggest that \textit{GPT} alone is insufficient for scenarios requiring highly accurate and project-specific guidance.

\vspace{-2em}
\subsubsection{RAG Answers}

\textit{RAG} answers were generally well-evaluated, with the study participants frequently highlighting their technical accuracy. By leveraging relevant documentation and source code, \textit{RAG} was able to provide answers aligned with Firefox's current artifacts. In several cases, evaluators preferred \textit{RAG} answers because they included concrete details and illustrative examples without unnecessary elaboration (Questions 5 and 6). 

\vspace{-0.5em}
\begin{mdframed}[frametitle=Question 5, subtitleaboveline=true, frametitlerule=true, subtitlebelowskip=0pt, subtitleaboveskip=5pt, skipabove=\baselineskip, skipbelow=\baselineskip, linecolor=black, linewidth=1.5pt]
  \footnotesize
  Can i convert nsString to nsCString?
  \mdfsubtitle{RAG's Answer:}
  Yes, you can convert an nsString (UTF-16) to an nsCString (UTF-8) using the $NS\_ConvertUTF16toUTF8$ helper class. For example, you can do nsCString $utf8String = NS\_ConvertUTF16toUTF8(utf16String);$ where utf16String is your nsString instance.
  \mdfsubtitle{Evaluator Comment:}
  [RAGAnswer] has the key function name and no other irrelevant information.
\end{mdframed} 
\vspace{-0.5em}

\begin{mdframed}[frametitle=Question 6, subtitleaboveline=true, frametitlerule=true, subtitlebelowskip=0pt, subtitleaboveskip=5pt, skipabove=\baselineskip, skipbelow=\baselineskip, linecolor=black, linewidth=1.5pt]
  \footnotesize
  Hi all, I'm working on making some changes to the new tab hover preview feature and I have come up against a strange issue. I am trying to set up a MutationObserver to listen to changes on other elements in the frontend (to decide if the tab preview should be disabled or not), but when I try to set it up (even with a no-op callback function), I get an $'Unexpected NS_ERROR_FAILURE'$ in the console. There is a javascript traceback pointing to my call, but there doesn't seem to be any way to get a more descriptive error message. Has anyone here run into something like this before?
  \mdfsubtitle{RAG's Answer:}
  It sounds like the $'Unexpected NS\_ERROR\_FAILURE'$ you're encountering might be due to the MutationObserver being set up on a node that isn't fully available or accessible at the time of the observer's initialization. Ensure that the target node exists and is fully loaded before setting up the observer. You can use $document.addEventListener($'DOMContentLoaded', ...$)$ or $window.onload$ to ensure the DOM is fully loaded before initializing the MutationObserver. Additionally, double-check that the node you're observing is not within a shadow DOM or another context that might restrict access.
  \mdfsubtitle{Evaluator Comment:}
  This [RAGAnswer] gives a small code example of what might work to fix the issue and also points out other possible issues.
\end{mdframed} 

Despite the positive aspects of \textit{RAG} answers, they were not flawless. While \textit{RAG} was more specific than \textit{GPT}, it occasionally missed key contextual details that \textit{Human} answers captured. Furthermore, \textit{RAG}'s reliance on retrieval sometimes led to overly detailed answers that, while accurate, could be overwhelming.

\subsection{RAG for Developer Assistance}

Results suggest that \textit{RAG} has potential for real-world use in large-scale projects like Mozilla Firefox. By integrating project-specific knowledge, \textit{RAG} could help developers answer questions requiring precise documentation references or technical details. An advantage is its ability to reduce the burden on senior developers, who often spend time addressing repetitive queries. Automating responses enables core maintainers to focus on high-impact development tasks rather than troubleshooting common issues.

Additionally, \textit{RAG} can improve onboarding by retrieving and synthesizing up-to-date information from evolving project documentation, reducing the learning curve for new contributors. However, successful deployment requires fine-tuning to ensure responses remain concise and contextually relevant. Prioritizing essential documentation while filtering out unnecessary details is critical for maintaining engagement in fast-paced discussions. While \textit{RAG} cannot replace human expertise, it presents a promising approach to enhancing productivity, improving information accessibility, and reducing response time in developer communities.

\vspace{-5px}
\section{Threats to Validity}
\label{sec:threats}

We discuss threats to the study's validity in what follows \cite{wohlin2012experimentation}.
\noindent
\textbf{Construct and Internal Validity.} Our study evaluates the effectiveness of Retrieval-Augmented Generation (\textit{RAG}) in the context of developer assistance within the Firefox project. A key threat to construct validity relates to our choice of the evaluation metrics, namely, helpfulness, comprehensiveness, and conciseness. While these attributes are relevant for assessing response quality, they may not fully capture all dimensions of usefulness in real-world development scenarios. Additional factors, such as accuracy, actionability, and developer trust in AI-generated responses, could influence adoption and effectiveness. \newcommand{\ConcisenessThreat}{{\color{\RevisionColor}Another concern is the requirement for one-paragraph responses, which may have systematically favored concise over comprehensive answers and potentially biased our evaluation since conciseness was one of the assessment criteria; however, this requirement was necessary \cite{chaves2021impact} to match communication expectations in developer chat channels.}}\ConcisenessThreat

Another concern stems from the selection of the questions analyzed. Although we applied systematic filtering criteria to extract a representative set of developers' questions, it is important to note that our dataset is inherently limited to Mozilla's Matrix chat rooms. Therefore, the sampling process excluded other relevant sources of developers' interaction, such as GitHub issues and mailing lists, which might vary in the characteristics of the questions and answers. Thus, we cannot generalize our findings to all sources of developers' online interaction, even for the Firefox project.


\newcommand{\ThreatsBiasChanges}{
{\color{\RevisionColor} The evaluation process may introduce bias, as all eight evaluators were Mozilla engineers. While their domain expertise ensures informed judgments, it also increases the risk of implicit preferences, particularly for \textit{human} answers. To mitigate this, all responses were anonymized and randomized, and the sources were unrevealed. However, one evaluator recognized their prior answer, as noted in the qualitative feedback. Although this was an isolated case, it highlights a limitation of our design. Future studies could involve external evaluators to mitigate this risk, although such an approach may introduce trade-offs related to domain expertise.
}
}
\ThreatsBiasChanges

\newcommand{\ThreatsVariaChanges}{
{\color{\RevisionColor} Our correlation analysis links preferred answers to those that are most helpful, comprehensive, or concise. We acknowledge that such distinctions may be less expressive when answers are similar in quality. 
Additionally, experts may have variability in their assessments, and we assessed agreement in four questions with overlapping evaluations. 
Our analysis focused on aggregate tendencies rather than individual-level modeling. Future work could adopt more sophisticated approaches, such as mixed-effects models, to better account for evaluators' variation.

} 
} 
\ThreatsVariaChanges

Ultimately, the models' performance is dependent on the specific LLM version used at the time of the study. GPT-4o and its \textit{RAG} variant are subject to continuous updates, which may affect their performance in future replications. It poses a reproducibility challenge, as different versions of the same model may yield significantly different results. While this is a known aspect in LLM research, we provide detailed methodological descriptions to aid in future replications. Besides, in the case of \textit{GPT}, OpenAI also provides the previous LLM versions in its API to mitigate this threat. However, we cannot assure that these versions will always remain available, as the \textit{GPT} models are proprietary.

\noindent
\textbf{Conclusion and External Validity.} The primary concern in terms of the external validity of our study is the generalizability of our findings. While Firefox is a widely used OSS project, different projects may have varying levels of documentation, distinct communication styles, and different developer expectations regarding AI-generated responses. Thus, while our results demonstrate the potential of \textit{RAG} in enhancing developer assistance, they may not be translated directly to other OSS communities without further validation.

Moreover, our study only considered responses from GPT-4o and GPT-4o enhanced with \textit{RAG}. The effectiveness of alternative LLMs or different retrieval-augmented approaches remains an open question. Variations in training data, retrieval mechanisms, or fine-tuning strategies could lead to different performance outcomes. Future research should explore a broader range of models and techniques to determine whether similar improvements can be observed across different AI systems.

Finally, while our evaluation provides insights into response quality, real-world adoption of RAG depends on additional factors such as usability and ethical considerations. LLMs can produce hallucinated or misleading answers, which could introduce risks in high-stake development tasks. Future work should explore long-term usability and developer trust to better understand how \textit{RAG}-based assistance tools can be effectively integrated into OSS development practices.

\vspace{-5px}
\section{Conclusion}
\label{sec:conclusion}

This paper evaluates the effectiveness of Retrieval-Augmented Generation (RAG) for LLMs in addressing technical queries from open source developers within the Mozilla Firefox project. Through systematic comparison of responses from human experts, baseline GPT models, and RAG-enhanced systems, we assessed the relative performance of these approaches in providing helpful, comprehensive, and concise technical assistance.

Our findings indicate that RAG-enhanced responses achieve greater comprehensiveness than human responses while maintaining comparable levels of helpfulness. This indicates that RAG systems can effectively leverage project-specific documentation and source code to generate more detailed technical explanations without sacrificing quality or relevance.

The study highlights the potential of RAG-based approaches to transform developer assistance in open source environments. By automating responses to common technical questions, preserving high response quality, such systems can substantially reduce the workload burden on core maintainers. This has important implications for the sustainability and scalability of open source projects, particularly as they grow in complexity and community size.

Several directions emerge for future research. First, we plan to extend our evaluation framework to additional open source projects across different domains to assess the generalizability of our findings and identify domain-specific factors that influence RAG effectiveness. Second, we aim to develop more sophisticated retrieval strategies that can handle specialized requirements such as system dependability, security considerations, and performance constraints. Third, longitudinal studies are necessary to understand the long-term adoption patterns of developers and the role of LLM-assisted knowledge systems in OSS development workflows. Finally, investigating the integration of RAG systems with existing developer tools and platforms, as well as evaluating them with end users, represents promising research avenues.

\section*{Acknowledgments}
This work was supported by the National Science Foundation grants 2236198, 2247929, 2303042 and 2303043. Also, CAPES: 88887.899310/2023-00, 88887.900069/2023-00 and 88887.915794/2023-00. CNPq: 140770/2021-6, 141180/2021-8, 140771/2021-2, 315711/2020-5, 141276/2020-7 and 141054/2019-0. CAPES/PROEX: 88887.373933/2019-00. CAPES/ PROCAD: 175956. FAPERJ: 200.510/2023, 211.033/2019 and 202.621/2019. Finally, we thank the Mozilla Firefox engineers for the collaboration in the experiments executions and answers analysis.

\balance
\bibliographystyle{ACM-Reference-Format}
\bibliography{references/refs}

\end{document}